\RequirePackage{fix-cm}
\RequirePackage{amsmath}
\documentclass[twocolumn,epjc3]{svjour3}  
\smartqed  
\RequirePackage{graphicx}
\RequirePackage{mathptmx}      
 
\usepackage{color}
\usepackage{mathrsfs}
\usepackage{amssymb, bm}
\usepackage{comment}
\usepackage[dvipsnames]{xcolor}
\usepackage{widetext}

\newcommand{\be}{\begin{equation}}
\newcommand{\ee}{\end{equation}}
\def\Box{\hbox{$\rlap{$\sqcup$}\sqcap$}}

\def\nn{\nonumber} 

\allowdisplaybreaks

\begin{document}

\title{Effective fluid mixture of tensor-multi-scalar gravity}

\author{Marcello Miranda  \thanksref{e1,addr1,addr2}
    \and
Pierre-Antoine Graham \thanksref{e2,addr3}
	\and
Valerio Faraoni  \thanksref{e3,addr4}
}

\thankstext{e1}{e-mail: marcello.miranda@unina.it}
\thankstext{e2}{e-mail: Pierre-Antoine.Graham@usherbrooke.ca}
\thankstext{e3}{e-mail: vfaraoni@ubishops.ca}


\institute{Scuola Superiore Meridionale, Largo San Marcellino 10, I-80138
Napoli, Italy\label{addr1} \and
INFN Sezione di Napoli, Complesso Universitario di Monte 
Sant’Angelo, Edificio G, Via Cinthia, I-80126 Napoli, Italy
\label{addr2} \and
Department of Physics, Universit\'e de Sherbrooke, 2500 
Boulevard de l'Universit\'e, Sherbrooke, Qu\'ebec, Canada J1K 2R1
\label{addr3} \and
Department of Physics \& Astronomy, Bishop's University, 
2600 College Street, Sherbrooke, Qu\'ebec, Canada J1M~1Z7 \label{addr4} 
}


\date{Received: date / Accepted: date}

\maketitle

\begin{abstract} 

We apply to tensor-multi-scalar gravity the effective fluid analysis based on the representation of the 
gravitational scalar field as a dissipative effective fluid. This 
generalization poses new challenges as the effective fluid is now a 
complicated mixture of individual fluids mutually coupled to each other 
and many reference frames are possible for its description. They are all 
legitimate, although not all convenient for specific problems, and they 
give rise to different physical interpretations. Two of these frames are 
highlighted.

\keywords{alternative theories of gravity \and tensor-multi-scalar gravity}

\end{abstract}

\section{Introduction}
\label{sec:1}
\setcounter{equation}{0}

It is well known that the scalar-tensor gravity field equations can be written as effective 
Einstein equations with an effective {\em dissipative} fluid in their 
right-hand side, built out of the Brans-Dicke-like scalar field $\phi$ 
present in 
the theory and of its first and second covariant derivatives 
\cite{Madsen:1988ph,Pimentel89,Faraoni:2018qdr,Quiros:2019gai}. The formalism has been generalized to ``viable'' Horndeski gravity 
\cite{Giusti:2021sku,Miranda:2022wkz,Giusti:2022tgq} and applied to Friedmann-Lema\^itre-Robertson-Walker 
(FLRW) cosmology \cite{Giardino:2022sdv}, to theories containing 
non-propagating scalar degrees of freedom \cite{Faraoni:2022doe,Miranda:2022brj}, and to 
specific scalar-tensor solutions \cite{Faraoni:2022jyd,Faraoni:2022fxo}. 
But what is the analogue of a multi-component fluid? Naturally, the simplest 
multi-fluid equivalent of a theory of gravity is tensor-multi-scalar 
gravity. Here we extend the effective fluid formalism to this class of 
theories. The task is much less obvious than it would appear at first sight 
because all the gravitational scalar fields couple to gravity, which 
makes them all couple to each other. In general, there can also be direct 
mutual couplings through their kinetic and potential terms in the action. 
In the presence of multiple {\em real} fluids decoupled from
each 
other, one can describe this mixture in the frame of an observer with 
timelike four-velocity $u^{\mu}$. This four-velocity can be that of the
comoving frame of one of the fluids, or it can be associated with any other 
observer. In general, it is difficult to define an average fluid 
\cite{EMMacC}. This 
means that the total stress-energy tensor $T_{\mu\nu}$ of the effective
fluid mixture, which is a tensor defined unambiguously, can be decomposed in 
many ways according to the four-velocity $u^{\mu}$ selected.  Each of these 
descriptions is legitimate but the description of the total mixture and its 
physical interpretation will depend on the observer $u^{\mu}$ selected to 
decompose $T_{\mu\nu}$. In particular, the density, pressure, heat flux
density, and anisotropic stresses of each fluid as ``seen'' from a 
particular observer $u^{\mu}$ will differ from those measured in the 
comoving frame of that fluid. 
To appreciate the difference between the descriptions of a fluid in 
different frames, consider a perfect fluid with four-velocity $u_{\mu}^*$ 
that, in its comoving frame, is described by the stress-energy 
tensor\footnote{We follow the notation and conventions of 
Ref.~\cite{Waldbook}: the metric 
signature is ${-}{+}{+}{+}$, $\kappa \equiv 8\pi G$, $G$ is Newton's 
constant, and units are used in which the speed of light $c$ is unity.}
\be
T_{\mu\nu}=\rho^* \, u^*_{\mu} \, u^*_{\nu} +P^* h^*_{\mu\nu} \,.
\ee
In the frame of  a different observer with timelike four-velocity $u^{\mu}$ 
related to $u^{\mu}_* $ by
\begin{eqnarray}
u_{\mu}^* &=& \gamma \left( u_{\mu}+v_{\mu} \right) \,,\\
&&\nonumber\\
\gamma &=& \frac{1}{\sqrt{ 1-v^2}}=-u^*_{\mu} u^{\mu}  \,,\\
&&\nonumber\\
v^2 &\equiv & v^{\alpha} v_{\alpha}>0 \,, \quad \quad v^{\alpha} 
u_{\alpha}=0 \,, \quad \quad 0 \leq v^2<1 \,,
\end{eqnarray}
this perfect fluid (now ``tilted'') will appear dissipative,  
with the different stress-energy tensor decomposition 
\cite{Maartens:1998xg,Clarkson:2003ts,Clarkson:2010uz} 
\be
T_{\mu\nu}=\rho  \, u_{\mu} \, u_{\nu} +P h_{\mu\nu} 
+q_{\mu} u_{\nu} +q_{\nu} u_{\mu} +\pi_{\mu\nu} \,, \label{imperfectfluid}
\ee
where \cite{Maartens:1998xg,Clarkson:2003ts,Clarkson:2010uz} 
\be
h_{\mu\nu}  \equiv g_{\mu\nu}+ u_{\mu} u_{\nu} \,,
\ee
\be
\rho = \rho^*  + \gamma^2 \,  v^2 \left( \rho^* + P^* \right)
= \gamma^2  \left( \rho^* + v^2 P^* \right) \label{sft1}
\ee
is the energy density, 
\be
P = P^*  +\frac{\gamma^2 \, v^2}{3}  \left( \rho^* + P^* \right)  
\label{sft2}
\ee
is the pressure, 
\be
q^{\mu} =  \left( 1+  \gamma^2 \, v^2 \right) \left( \rho^* + P^* 
\right) v^{\mu}= \gamma^2  \left( \rho^* + P^* 
\right) v^{\mu}   \label{sft3}
\ee
is the energy flux density, and 
\be
\pi^{\mu\nu} = \gamma^2 \left( \rho^* + P^* \right) \left( v^{\mu} \, 
v^{\nu} -\frac{v^2}{3} \, h^{\mu\nu} \right) \label{sft4} 
\ee
is the anisotropic stress tensor. 
It is clear that the (spatial) vector $q^{\mu}$ arises 
solely due to the relative motion between the two frames, {\em i.e.}, to 
the (spatial) vector $ v^{\mu}$. In this context it is problematic to 
interpret this purely convective current as a heat flux according to Eckart's 
generalization of Fourier's law \cite{Eckart40} 
\be 
q_{\mu}= -K \left( 
h_{\mu\nu} \nabla^{\nu}T + T \, \dot{u}_{\mu} \right) \,, 
\label{generalizedFourier} 
\ee 
where $T$ is the temperature and $K$ is the 
thermal conductivity. This law expresses the fact that heat conduction is
caused not only by spatial temperature gradients but also by an 
``inertial'' contribution due to the fluid acceleration \cite{Eckart40}.  

The situation becomes more complicated when multiple fluids are coupled to 
each other and even more when they are {\em effective} fluids and they all 
couple explicitly with the curvature\footnote{We do not consider 
derivative couplings in this work.} (more precisely, with the Ricci scalar 
$R$) and to each other, which is the situation in tensor-multi-scalar 
gravity. In this work we discuss two possibilities, but other frames may 
be more 
convenient for specific problems.

Rather surprisingly, in tensor-multi-scalar gravity formulated in the Jordan 
conformal frame, one can obtain a particular frame as a sort of fictitious 
``average'' frame, which is generally not possible with real fluids 
\cite{EMMacC}. It 
is obtained by identifying the coupling function of the scalars to $R$ 
(which depends on {\em all} the scalar fields in the theory) with a new 
field $\psi$ and amounts to a redefinition of the scalar fields. This 
procedure is routine in tensor-single-scalar gravity, in which the only 
Brans-Dicke-like field is redefined for convenience, without much 
consequence or interpretation. In tensor-multi-scalar gravity, instead, this 
redefinition takes a new meaning. It identifies a four-velocity and a sort 
of ``average'' frame because there is only one Ricci scalar $R$ and all the 
scalar fields in the theory couple to it. This ingredient is missing for 
real fluids, which do not couple to the curvature and have no ``average'' 
frame \cite{EMMacC}.

In the following we analyze tensor-multi-scalar gravity in its Jordan 
(conformal) frame formulation. It is possible to discuss it from the point 
of view of the ``average'' observer, or from the comoving frame of each 
fluid, or from that of any other timelike observer $u^{\mu}$. It is 
important to 
remember that these descriptions will be different and will provide 
different physical interpretations of the mechanical and thermal aspects of 
the fluid mixture, and that these are all legitimate (hence one should not 
strive to identify the ``correct'' one). The point is that some of these 
formulations (originating different decompositions of the total $T_{\mu\nu}$ 
based on different $u^{\alpha}$) will be more convenient, and some others 
will be less convenient, for specific physical problems. One should adopt
the formulation that is most convenient for the particular problem at hand 
without prejudice. For example, analyses of the quark-gluon plasma created 
in heavy ion collisions universally employ the Landau (or energy) frame 
\cite{BRAHMS:2004adc,PHOBOS:2004zne,STAR:2005gfr,PHENIX:2004vcz,Monnai:2019jkc} 
in which there is no heat flux\footnote{This frame is found to be non-unique in Ref.~\cite{Romero-Munoz:2014foa}.}
while in FLRW cosmology, where comoving coordinates are the standard, 
relativistic fluids are routinely described in their comoving (or Eckart) 
frame \cite{Waldbook,EMMacC}.
  
Here we are interested in the fluid-mechanical equivalent and in 
the thermal description of tensor-multi-scalar 
gravity, where the fluids in the mixture are {\em effective} fluids and they 
all couple explicitly with $R$ and with each other. This is a very specific 
situation and our choices, although convenient in this problem, are not 
meant to be recipes with universal convenience (although aspects of our 
discussion may apply to other situations as well). After this discussion, we 
present an alternative view of the first-order thermodynamics of 
tensor-multi-scalar gravity in the Einstein conformal frame, while the last 
section summarizes our conclusions.

\section{Tensor-multi-scalar gravity in the Jordan conformal frame}
\label{sec:2}

Let us begin with a convenient Jordan frame formulation of 
tensor-multi-scalar gravity (without 
derivative couplings). We adopt most of the notations specific to 
tensor-multi-scalar gravity 
used in Ref.~\cite{Hohmann:2016yfd}. There are $N$ scalar fields of 
gravitational nature $\left\{ \phi^A 
\right\}$, with $A=1,2, \, ... 
\, , N$,  all coupled nonminimally with the Ricci scalar $R$ and between 
themselves, as described by the action
\begin{eqnarray}
S_\mathrm{TMS} &=& \frac{1}{2\kappa} \int d^4x \sqrt{-g} \left[ F( \phi^J) 
R 
-Z_{AB} (\phi^J)  g^{\mu\nu} \nabla_{\mu} \phi^A \nabla_{\nu} \phi^B 
\right. \nonumber\\
&&\nonumber\\
&\, & \left. - V( 
\phi^J) \right] + S^\mathrm{(m)} \,,\label{action}
\end{eqnarray}
where capital indices $A,B, J \,, ... $ label the 
scalar  fields in the multiplet $\left\{ \phi^1 \,, ... \,, \phi^N 
\right\}$, $g$ is the determinant of the spacetime metric $g_{\mu\nu}$, 
$\nabla_{\mu}$ is the associated covariant derivative, and $V$ is a scalar 
field potential. The Einstein summation convention is used also on the 
multiplet indices $J$. The coupling 
function $F \left( \phi^1 \,, ... \,, \phi^N \right) $ depends on all the 
$ \phi^A $, {\em i.e.}, $ \partial F/\partial \phi^I \neq 0 \; \forall 
I\in \left\{1, \, ... \,, N \right\}$, or else some of the scalar fields 
would not be coupled 
directly to $R$ and would lose their status of gravitational scalar 
fields.\footnote{The nature of these scalar fields (gravitational or not) 
depends on the conformal frame \cite{Sotiriou:2007zu}. Here we refer to 
the Jordan conformal frame.} $F$ is assumed to 
be positive to keep the 
effective gravitational coupling $G_\mathrm{eff} \simeq 1/F$ positive. 

The matrix $Z_{AB} \left( \phi^1 \,, ... \,, \phi^N \right) $ acts as a 
Riemannian metric on the scalar field space of coordinates $\left\{ \phi^1 
\,, ... \,, \phi^N \right\}$. $Z_{AB}$ can be taken to be symmetric 
without 
loss of generality because it multiplies the combination of kinetic terms 
$\nabla^{\alpha} \phi^A \nabla_{\alpha} \phi^B$ symmetric in $A$ and $B$. 
The elements of $Z_{AB}$ are all positive to avoid introducing unstable 
phantom fields. In general, also the potential $ V\left( \phi^1 \,, ... 
\,, 
\phi^N \right) $ depends on multiple fields (although it is not important 
that it depends on {\em all} these fields, which is instead crucial for 
the coupling function $F$).

Since the matrix $Z_{AB}$ is real and symmetric, it can be diagonalized at 
each spacetime point $x^{\mu}$ and has positive eigenvalues, turning the 
sum of kinetic terms appearing  in the action~(\ref{action}) into 
\begin{eqnarray}
 Z_{AB} \left( \phi^J \right)  g^{\mu\nu} \nabla_{\mu} \phi^A 
\nabla_{\nu} \phi^B &=&
\bar{Z}_{AB} \left( \bar{\phi}^J \right)   g^{\mu\nu} 
\nabla_{\mu}  \bar{\phi}^A \nabla_{\nu} \bar{\phi}^A \nonumber\\
&&\nonumber\\
&=& \sum_{A=1}^N    \bar{Z}_{A} \left( \bar{\phi}^J \right) g^{\mu\nu} \,  
\nabla_{\mu}  \bar{\phi}^A \nabla_{\nu} \bar{\phi}^A  \,,\nonumber\\
&&
\end{eqnarray}
where a bar denotes fields in the system of principal axes of the matrix 
$Z_{AB}$ in field space, and 
\be
\bar{Z}_{AB} = \mbox{diag} \left( \bar{Z}_1, \, ... \, , \bar{Z}_N \right) 
\ee 
is the diagonal form of 
$Z_{AB}$. This diagonalization, however, is not crucial and we will not 
use it explicitly, retaining the non-diagonal form of $Z_{AB}$ in 
our formulae.

\section{Multi-fluid decomposition}
\label{sec:3}

The total stress-energy tensor is obtained by varying the action~(\ref{action}) 
with respect to $g^{\mu\nu}$. Using $\partial_{A}\equiv\partial/\partial\phi^A$ and $D_{AB} \equiv Z_{AB} + \partial_{AB} F$, the associated equation of motion reads
\begin{align}
    G_{\mu \nu} = \kappa \, T_{\mu \nu} +\frac{\kappa\,T^\mathrm{(m)}_{\mu\nu}}{F}
    \label{Teff_0} \,,
\end{align} 
where $G_{\mu \nu} \equiv R_{\mu \nu}-\frac{1}{2} \, g_{\mu \nu} R$ is the Einstein tensor, $T^\mathrm{(m)}_{\mu\nu}\equiv-\tfrac{2}{\sqrt{-g}} \, \frac{ \delta  S^\mathrm{(m)}}{ \delta g^{\mu\nu} }$ is the matter stress-energy tensor 
and 
\begin{align}
\kappa \, T_{\mu \nu} &= \dfrac{1}{F} \, 
\partial_{A} F \left[\nabla_\mu \nabla_\nu \phi^A - g_{\mu \nu} \square 
\phi^A\right] + \dfrac{D_{AB}}{F} \, \nabla_\mu \phi^A \nabla_\nu \phi^B \nonumber \\
&\nn\\
    &- \dfrac{1}{2F} \left(Z_{AB} +2\partial_{AB} F\right) g_{\mu \nu} \nabla_\rho \phi^A \nabla^\rho \phi^B - \dfrac{ V}{2F} \, g_{\mu \nu} \,. \label{Teff_1}
\end{align}
The equation of motion obtained by variation of the action with respect to $\phi^A$
reads 
\begin{align}
2 Z_{AB} \Box\phi ^{A  }=&\,+\partial _{B }{}F\, R - \partial _{B }{}V_{  }- \partial _{B}{}Z_{AC  } \nabla _{\alpha}\phi ^{A  } \nabla ^{\alpha}\phi ^{C  }
    &\nn\\
    &\nn\\
    & + 2 \partial _{A }{}Z_{  C  B  } \nabla _{\alpha}\phi ^{C  } \nabla ^{\alpha}\phi ^{A  } \,.\label{boxphi_1}
\end{align}
We can obtain the expression of the Ricci scalar from \eqref{Teff_0},
\begin{align*}
    R =\,& \frac{2 V}{F} + \frac{3 \partial _{A}F}{F} \, \Box\phi ^{A }-\frac{\kappa\,T^\mathrm{(m)}}{F} + \frac{\big(3\partial _{AB }F_{}+ Z_{A B}\big)}{F} \, \nabla _\alpha\phi ^{B  } \nabla ^\alpha\phi ^{A  } \,,
\end{align*}
where $T^{(m)} \equiv g^{\mu\nu}T^{(m)}_{\mu\nu}$ is the trace of the matter stress-energy tensor.
With this expression, Eq.~\eqref{boxphi_1} turns into
\begin{align*}
    0=\,&\frac{(3 \partial _{A}{}F_{} \partial _{B }{}F_{} + 2 F Z_{A B }) }{F} \, \Box\phi ^{A }+\frac{2 \partial _{B }{}F_{ }\, V}{F} - \partial _{B }{}V_{ } -\frac{\partial_{B}F}{F}\kappa\,T^\mathrm{(m)} \nn\\
    &\nn\\
    &+ \frac{\partial _{B }{}F_{ }  }{F}(3 \partial _{AC}F + Z_{A C }) \nabla _{\alpha}\phi ^{A } \nabla ^{\alpha}\phi ^{C } \nn\\
    &\nn\\
    &+ \ (2 \partial _{A }{}Z_{ B C } - \partial _{B}{}Z_{  A C }) \nabla _{\alpha}\phi ^{A } \nabla ^{\alpha}\phi ^{C } \,.
\end{align*}
Assuming $\mathrm{det}\left( 3 \partial _{A}{}F_{} \partial _{B }{}F_{} + 2 F Z_{A B } \right)\neq0$, we use the matrix $M^{AB}\equiv \left( 3 \partial _{A}{}F_{} \partial _{B }{}F_{} + 2 F Z_{A B } \right)^{-1}$ to isolate $\Box\phi^{A}$, obtaining
\begin{align}
    \Box\phi^A=\,&M^{AB}\Big[ F\,\partial _{B }{}V_{ } -2 V\,\partial _{B }{}F+\kappa\,T^\mathrm{(m)}\,\partial_{B}F \nn\\
    &\nn\\
    &- \partial _{B }{}F_{ } (3 \partial _{AC}F + Z_{A C }) \nabla _{\alpha}\phi ^{A } \nabla ^{\alpha}\phi ^{C } \nn\\
    &\nn\\
    &- \ F(2 \partial _{A }{}Z_{ B C } - \partial _{B}{}Z_{  A C }) \nabla _{\alpha}\phi ^{A } \nabla ^{\alpha}\phi ^{C }\Big]\,.
\end{align}
The goal of the decomposition given here is to separate $T^{\mu \nu}$ so that each part can be decomposed in the frame of a given fluid. Each fluid then receives an individual stress-energy tensor contribution. The number of purely convective terms is minimised by such a decomposition to allow for a clearer description of the intrinsic dissipative properties of each fluid. 

Assuming the gradient of each scalar field to be timelike,  
\begin{align}
    X^{A} \equiv -\frac{1}{2} \, \nabla^{\mu} \phi^A \nabla_{\mu} \phi^A > 0 \,, 
\end{align}
we define the $\phi^A$-fluid four-velocity
\begin{align}
    u_{\mu}^{A} \equiv  \dfrac{\nabla_{\mu}\phi^A}{\sqrt{2X^{A}}} \,. 
\label{phi_4velocities}
\end{align} 
At this point, in order to avoid ambiguities, all the multiplet summations in this section will be  written with an explicit summation symbol.
The above identification between a scalar field and an associated effective fluid allows us to rewrite the scalar field derivatives in term of kinematic quantities \cite{Miranda:2022wkz}.
The second derivative $\nabla_{\mu}\nabla_{\nu}\phi^A$ in Eq.~\eqref{Teff_1} can be expanded as 
\begin{align}
    \nabla_{\mu}\nabla_{\nu}\phi^A=&\,\sqrt{2X^{A} }\left(\sigma^{A}{} _{ \mu\nu}+\frac{1}{3}\, \Theta^{A}{}  h^{A}_{ \mu\nu}-2\dot{u}^{A}{} _{(\mu}u^{A}{} _{\nu)}\right)\nn\\
    &\nn\\
    &\,- \left(\Box\phi^{A}-\sqrt{2X^A}\,\Theta^{A}\right) \, u^{A}{} _{\mu}u^{A}{} _{\nu}\,,
\end{align}
where  $h^{A}{}_{\mu\nu} \equiv g_{\mu\nu}+u^{A}{}_{\mu}u^{A}{}_{\nu}\,$ is the three-metric of the hypersurface orthogonal to the four-vector $u^{A}{}_\mu$, $\Theta^A \equiv \nabla_\mu u^{A}{}^{\mu}$ is the expansion tensor associated with the $A$-fluid, and $\sigma^{A}{}_{\mu\nu}\equiv \tfrac{1}{2} \left( h^{A}{}_{ac}\nabla^{c}u^{A}{}_{b} 
+h^A{}_{bc}\nabla^{c}u^A{}_{a} \right) -\tfrac{1}{3} \, \Theta^{A} \, h^{A}_{ab}$.

With this result, Eq.~\eqref{Teff_1} becomes
\begin{align}
    \kappa \, T_{\mu \nu}= \,&\sum_{A,B}\Bigg\{\frac{1}{F}\left(2\sqrt{X^{A}X^{B}} D_{AB}+\sqrt{2X^A}\partial_{A}F\,\Theta^{A}\delta_{AB} \right)u^A{}_{\mu} u^{B}{}_\nu\nn\\
    &\nn\\
    &\,+\frac{\partial_{B}F}{F}\delta_{AB}\left(-\Box\phi^A+  \frac{\sqrt{2X^A}}{3}\Theta^{A}\right)h^{A}{}_{\mu\nu}\nn\\
    &\nn\\
    &\,-\frac{1}{2F}\left[2\sqrt{X^{A}X^{B}} \left(Z_{AB} +2\partial_{AB} F\right)\, u^A{}_{\rho} u^{B}{}^\rho \right]g_{\mu \nu}\nn\\
    &\nn\\
    &\,+\dfrac{\sqrt{2X^A}\partial_{B} F}{F} \,\delta_{AB} \Big(\sigma^{A}{} _{ \mu\nu}-2\dot{u}^{A}{} _{(\mu}u^{A}{} _{\nu)}\Big)\Bigg\}-\frac{V}{2F}g_{\mu\nu}\,, \label{Teff_2}
\end{align}
where
\begin{align}
    \Box\phi^A=\,&\sum_{A,B}\bigg\{M^{AB}\Big[ F\,\partial _{B }{}V_{ } -2 V\,\partial _{B }{}F+\kappa\,T^\mathrm{(m)}\,\partial_{B}F \nn\\
    &\nn\\
    &- 2\sum_{C}\Big(\sqrt{X^AX^C}\partial _{B }{}F_{ } \left(3 \partial _{AC}F + Z_{A C }\right) u ^{A }{}_\alpha u^{C }{}^{\alpha} \nn\\
    &\nn\\
    &+\sqrt{X^AX^C} \, F\left(2 \partial _{A }{}Z_{ B C } - \partial _{B}{}Z_{  A C }\right) u ^{A }{}_\alpha u^{C }{}^{\alpha}\Big)\Big]\bigg\}\,. \label{boxphi}
\end{align}
Since this equation does not depend on four-velocity gradients, we can interpret it as a purely inviscid contribution to the stress-energy tensor mixture.

If we rewrite the metric as
\begin{align}
    g_{\mu\nu}=\,&h^{A}{}_{\mu\nu}-u^{A}{}_{\mu}u^{A}{}_{\nu}=\frac{1}{N}\, \sum_{J}\left(h^{J}{}_{\mu\nu}-u^{J}{}_{\mu}u^{J}{}_{\nu}\right)\nn\\
    =\,&\frac{1}{N} \, \sum_{A,B}\delta_{AB}\left(h^{B}{}_{\mu\nu}-u^{A}{}_{\mu}u^{B}{}_{\nu}\right)
\end{align}
and we define
\begin{align}
    \mathcal{T}\equiv -\frac{1}{N} \, \sum_{A,B}\left\{\sqrt{X^{A}X^{B}} \left(Z_{AB} +2\partial_{AB} F\right)\, u^A{}_{\rho} u^{B}{}^\rho \right\}-\frac{V}{2N}
\end{align}
then, writing explicitly the summations, the stress-energy tensor assumes the form
\begin{widetext}
\begin{align}
    \kappa \, T_{\mu \nu}=\,&\sum_{A,B}\Bigg\{\frac{1}{F}\left(2\sqrt{X^{A}X^{B}} \,  D_{AB}+\sqrt{2X^A} \, \partial_{A}F\,\Theta^{A}\delta_{AB} -\mathcal{T}\delta_{AB} \right)u^A{}_{\mu} u^{B}{}_\nu+\frac{\partial_{B}F}{F}\,\delta_{AB}\left(\mathcal{T}-\Box\phi^B+  \frac{\sqrt{2X^B}}{3} \, \Theta^{B}\right)h^{A}{}_{\mu\nu}\nn\\
    &\nn\\
    &\,+\dfrac{\sqrt{2X^B}\partial_{B} F}{F} \delta_{AB} \, \Big(\sigma^{B}{} _{ \mu\nu}-2\dot{u}^{A}{} _{(\mu}u^{B}{} _{\nu)}\Big)\Bigg\} \,,\label{cross_terms}
\end{align}
\end{widetext}
which is interpreted as a mixture of interacting imperfect fluids.

\section{``Average'' or ``$\psi$-'' description}
\label{sec:4}

Let us discuss another possible procedure. In the following we  redefine the fields $\phi^A$ 
but, before 
proceeding, it is essential to note (and remember through the rest of 
this work) that {\em all} these fields couple directly with the Ricci 
scalar $R$ 
through $F$ and they all play a role of in determining the properties of 
the effective fluid equivalent to the tensor-multi-scalar theory and the 
effective gravitational coupling $G_\mathrm{eff} \equiv F^{-1}$. (Their 
role may be different as, in general,  $F\left(\phi^1, \, ... \, , \phi^N 
\right)$ is not symmetric in all its arguments.)  In 
particular, the effective temperature of this multi-component fluid is 
determined by {\em all} the fields $\phi^A$ and the upcoming redefinition 
of these fields does not change this fact. 

We  proceed to redefine the scalar field 
multiplet as in Ref.~\cite{Hohmann:2016yfd}, which is standard practice 
in single-scalar-tensor gravity. We can rename the coupling 
function by electing it to be a Brans-Dicke-like scalar,
\be
\psi \equiv F\left( \phi^1 \,, ... \,, \phi^N \right) \,,
\ee
and we then have the $N$ scalar fields $\left\{ \psi, \phi^1, \, ... 
\,, \phi^{N-1} \right\}$. This mathematically convenient procedure 
effectively makes only the field $\psi$ couple explicitly to $R$  but the 
reader should not be fooled into believing that the remaining fields 
$\phi^A$ do not couple to gravity. In fact, all the fields 
$ \phi^A $ are coupled to $\psi$ (and also to each other), which makes 
them couple also to gravity. Indeed, they were explicitly coupled to 
gravity before the field redefinition $\psi \equiv F$ and the physics   
does not change. The action~(\ref{action}) is recast as 
\cite{Hohmann:2016yfd} 
\begin{eqnarray}
S_\mathrm{TMS} &=& \frac{1}{2\kappa} \int d^4x \sqrt{-g} \bigg[ \psi R 
-\frac{\omega}{\psi} \,\nabla^{\alpha} \psi 
\nabla_{\alpha}\psi  
 \nonumber\\
&&\nonumber\\
&\, &  -Z_{AB}  \nabla^{\mu} \phi^A \nabla_{\mu} 
\phi^B  -V  \bigg] 
+ S^\mathrm{(m)}  \,,\label{action2}
\end{eqnarray}
where 
\begin{eqnarray}
\omega &=& \omega \left( \psi, \phi^A \right) \,, \quad \quad 
2\omega+3>0 \,, \\
&&\nonumber\\
Z_{AB} &=& Z_{AB} \left( \psi, \phi^J \right) >0 \,,\\
&&\nonumber\\
V &=& V\left( \psi, \phi^A \right) \,.
\end{eqnarray}

The field equations for $g_{\mu\nu}, \psi, \phi^A$ obtained by 
varying the action~(\ref{action2}) are \cite{Hohmann:2016yfd}  
\begin{align}
    G_{ \mu\nu}=\,&
    \frac{1}{\psi}\left( \nabla _{\mu} \nabla_{\nu}\psi-g_{ \mu\nu}\, \Box\psi \right)\nn\\
    &\nn\\
    &+\frac{\omega}{\psi^2}\left(  \nabla_{\mu}\psi  \nabla_{\nu}\psi 
    - \frac{1}{2} \, g_{ \mu\nu}   \nabla_{\alpha}\psi  \nabla^{\alpha}\psi \right)\nn\\
    &\nn\\
    &+\frac{Z_{AB}}{\psi} \left( \nabla_{\mu}\phi^A \nabla_{\nu}\phi^B 
    - \frac{1}{2} \, g_{\mu\nu}  \nabla_{\alpha}\phi^A \nabla^{\alpha} \phi^B \right)\nn\\
    &\nn\\
    &- \frac{g_{\mu\nu} V}{2 \psi } +\frac{\kappa}{\psi } \, T_{ \mu\nu}^\mathrm{(m)}\,,\label{fe1}\\
    &\nn\\
    Z_{AB} \, \Box \phi^B=\,&
    \left( \frac{1}{2} \, \partial_A {}Z_{BC} - \partial_B {}Z_{AC}\right) \nabla_{\alpha} \phi^C \nabla^{\alpha} \phi^B \nn\\
    &\nn\\
    &+ \frac{1}{2\psi} \, \partial_A \omega \, \nabla_{\alpha}\psi  \nabla^{\alpha}\psi  - \partial_{\psi }{}Z_{AB}\,   \nabla_{\alpha}\psi  \nabla^{\alpha}\phi^B \nn\\
    &\nn\\
    &+\frac{1}{2}\partial_A. V\,, \label{fe2}\\
    &\nn\\
    \Box\psi  =\,&\frac{\psi}{2\omega} \left( \partial_{\psi }{}V   + \partial_{\psi }{}Z_{AB}\,  \nabla_{\alpha}\phi^B \nabla^{\alpha}\phi^A  - R\right)\nn\\
    &\nn\\
    & - \frac{\partial _{A }{}\omega}{\omega } \nabla _{\alpha}\psi  \nabla ^{\alpha}\phi ^{A  } + \frac{(\omega  - \psi  \partial _{\psi }{}\omega )}{2 \psi  \omega } \nabla _{\alpha}\psi  \nabla ^{\alpha}\psi\,, \label{fe3}
\end{align}
where we have used the notation $\partial_{A}\equiv\partial/\partial\phi^{A}$
and $\partial_{\psi}\equiv\partial/\partial\psi$.

Using the metric field equations we can express the Ricci scalar in terms of the matter and effective stress-energy tensors,
\begin{align}
    R =  \,& \frac{3 }{\psi } \, \Box\psi   + \frac{ \omega }{\psi ^2} \, \nabla _{\alpha}\psi  \nabla ^{\alpha}\psi\nn\\
    &\nn\\
    &+ \frac{Z_{AB  }}{\psi } \, \nabla _{\alpha}\phi^A \nabla^{\alpha}\phi ^B + \frac{2 V}{\psi }- \frac{\kappa}{\psi } \, T^{\mathrm{(m)}}
\end{align}
where $T^\mathrm{(m)} \equiv g^{ \mu\nu}T_{ \mu\nu}^\mathrm{(m)}$. Then, the equation of motion for $\psi$ turns into
\begin{align}
    \Box\psi  = \,& \frac{1}{3 + 2 \omega }\Big[ \left ( \psi \partial _{\psi }{}Z_{AB }  - Z_{AB }\right) \nabla _{\alpha}\phi ^{A  } \nabla ^{\alpha}\phi ^{B  }\nn\\
    &\nn\\
    &- 2 \partial _{A }{}\omega \nabla _{\alpha}\psi  \nabla ^{\alpha}\phi ^{A  } 
    - \partial _{\psi }{}\omega  \nabla _{\alpha}\psi  \nabla ^{\alpha}\psi\nn\\
    &\nn\\
    &+\psi\partial _{\psi }{}V - 2 V +\kappa\,T^\mathrm{(m)}\Big] \,.
\end{align}
Finally, we define the effective stress-energy tensor as
\begin{align}
    \kappa \,T_{ \mu\nu}\equiv\,&
    \frac{1}{\psi}\left(\nabla _{\mu}\nabla _{\nu}\psi-g_{ \mu\nu}\, \Box\psi\right)\nn\\
    &\nn\\
    &+\frac{\omega}{\psi^2}\left(  \nabla _{\mu}\psi  \nabla _{\nu}\psi 
    - \frac{1}{2} \, g_{ \mu\nu}   \nabla _{\alpha}\psi  \nabla ^{\alpha}\psi \right)\nn\\
    &\nn\\
    &+\frac{Z_{AB}}{\psi} \left( \nabla _{\mu}\phi ^{A  } \nabla _{\nu}\phi ^{B  }
    - \frac{1}{2}g_{ \mu\nu}  \, \nabla _{\alpha}\phi ^{A  } \nabla ^{\alpha}\phi ^{B  }\right)\nn\\
    &\nn\\
    &- \frac{g_{ \mu\nu} V}{2 \psi } \,.
\end{align}
We can now move to the effective fluid picture.

\section{Comoving (Eckart)  frame of $\psi$-fluid} 
\label{sec:5}

Assume that the gradient of $\psi$ is timelike; using
\be
X \equiv -\frac{1}{2} \, \nabla^\mu \psi 
\nabla_{\mu} \psi >0 
\ee
we define the effective fluid four-velocity
\be
u^{\mu} = \frac{ \nabla^\mu \psi }{\sqrt{2X}} \label{4-velocity}
\ee
which is normalized, $u^{\mu} u_{\mu} =-1$ (but the sign of the right-hand 
side of this definition must be adjusted to keep $u^{\mu}$ a 
future-oriented vector, which is crucial in discussions of dissipation  
which is time-irreversible). 
In general, the $\phi^A$-fluids are tilted with respect to the 
$\psi$-fluid, {\em i.e.}, $u^{A}{}^{\mu}$ and $u^{\mu}$ have different 
directions. Using the $u^{\mu}$ of the 
$\psi$-fluid we perform the usual $3+1$ splitting of spacetime into the 
time direction and the 3-space ``seen'' by the observer with four-velocity 
$u^{\mu}$. This 3-space has Riemannian metric
\be
h_{\mu\nu} \equiv g_{\mu\nu}+u_{\mu}u_{\nu} \,.\label{h}
\ee
The kinematic quantities (expansion tensor $\Theta_{\mu\nu}$, expansion 
scalar $\Theta=\nabla_{\mu} u^{\mu}$, shear tensor $\sigma_{\mu\nu}$, 
shear scalar, and acceleration $\dot{u}^{\mu}$)
associated with $u^{\mu}$ are the same as 
those calculated for single-scalar-tensor gravity 
in~\cite{Faraoni:2018qdr}.  In fact, 
their definitions are purely kinematic and theory-independent since they 
do not use the field equations but only the definition~(\ref{4-velocity}) 
of $u^{\mu}$. These kinematic quantities are 
straightforward, although lengthy to compute. Since they are used here, we report them in~\ref{sec:appendix}. 
 
The field equations~(\ref{fe1}) have the form of effective Einstein 
equations with an effective stress-energy tensor in their right-hand side, 
which can be seen as the stress-energy tensor of a {\em dissipative} 
multi-component fluid of the form
\be
T_{\mu\nu}  =\left( P+\rho \right)u_{\mu}u_{\nu} 
+Pg_{\mu\nu} 
+q_{\mu}u_{\nu}+q_{\nu}u_{\mu} +\pi_{\mu\nu} 
\ee
where 
\be
\rho = T_{\mu\nu}  u^{\mu} u^{\nu} 
\ee
is the effective energy density,
\be
q_{\mu} = -T_{\alpha\beta}  u^{\alpha} {h_{\mu}}^{\beta} 
\ee
is the effective heat current density describing heat conduction,
\be
\Pi_{\alpha\beta} = P h_{\alpha\beta} +\pi_{\alpha\beta} 
=T_{\mu\nu}   
{h_{\alpha}}^{\mu} {h_{\beta}}^{\nu} 
\ee
is the effective stress tensor,
\be
P =\frac{1}{3} \, g^{\alpha\beta} \Pi_{\alpha\beta} = \frac{1}{3} \, 
h^{\alpha\beta} T_{\alpha\beta}  
\ee
is the effective isotropic pressure, and the trace-free part of the stress 
tensor
\be  
\pi_{\alpha\beta} = \Pi_{\alpha\beta}-P h_{\alpha\beta}
\ee
is the effective anisotropic stress tensor. $q^{\mu}$, 
$\Pi_{\alpha\beta}$, and  $\pi_{\alpha\beta}$ are purely spatial with 
respect to $u^{\mu}$. 
The fluid description is obtained by expressing the
derivatives of $\psi$ in terms of the relative effective fluid
four-velocity~(\ref{4-velocity}) and kinematic quantities,
\begin{align}
    \nabla_{\mu}\psi=&\,\sqrt{2X }\,u _{\mu}\,,\\
    &\nn\\
    &\nn\\
    \nabla_{\mu}X =&\,-\dot{X} \,u _{\mu}+h _{ \mu\nu}\nabla^{\nu}X=-\dot{X} \,u _{\mu}-2X \,\dot{u} _{\mu}\,,\\
    &\nn\\
    \nabla_{\mu}\nabla_{\nu}\psi=&\,\nabla_{\mu}\left(\sqrt{2X }\,u _{\nu}\right)\nn\\
    &\nn\\
    =&\,\sqrt{2X } \, \nabla_{\mu}u _{\nu}-\frac{\dot{X} }{\sqrt{2X }} \, u _{\mu}u _{\nu}-\sqrt{2X}\dot{u}_{\mu}u_{\nu}\nn\\
    &\nn\\
    =&\,\sqrt{2X }\left(\sigma _{ \mu\nu}+\frac{1}{3}\Theta  h _{ \mu\nu}-2\dot{u} _{(\mu}u _{\nu)}\right)-\frac{\dot{X} }{\sqrt{2X }} \, u _{\mu}u _{\nu}\,.
\end{align}
Furthermore, we have 
\begin{align}
    \Box\psi=\sqrt{2X } \, \Theta +\frac{\dot{X} }{\sqrt{2X }} \,,
\end{align}
therefore the $\psi$-equation of motion reads
\begin{align}
    \frac{\dot{X}}{\sqrt{2X}}  = \,& -\sqrt{2X}\, \Theta \nn\\
    &\nn\\
    &+\frac{1}{3 + 2 \omega }\Big[ \left ( \psi \partial _{\psi }{}Z_{AB }  - Z_{AB }\right) \nabla _{\alpha}\phi ^{A  } \nabla ^{\alpha}\phi ^{B  }\nn\\
    &\nn\\
    &- 2\sqrt{2X } \partial _{A }{}\omega\, u _{\alpha}  \nabla ^{\alpha}\phi ^{A  } 
    +2 \partial _{\psi }{}\omega  X \nn\\\
    &\nn\\
    &+\psi\partial _{\psi }{}V - 2 V +\kappa\,T^\mathrm{(m)}\Big] \,.
\end{align}
We need these equations to eliminate the dependence of $T_{ \mu\nu}$ on $\dot{X}$ and on $\Box\psi$. Indeed, prior to using the equation of motion for $\psi$, one obtains
\begin{align}
    \kappa T_{ \mu\nu}=\,&\left(\frac{ V}{2 \psi }+ \frac{ X {} \omega }{\psi ^2} + \frac{\sqrt{2X}}{ \psi }\Theta \right)u_{\mu}u_{\nu}\nn\\
    &\nn\\
    &+\left(- \frac{ V}{2 \psi }+ \frac{ X {} \omega }{\psi ^2}- \frac{\dot{X} {}}{ \sqrt{2X} \psi }- \frac{2 \sqrt{2X}   }{3 \psi }\Theta  \right)h_{ \mu\nu}\nn\\
    &\nn\\
    &- 2\frac{\sqrt{2X}  }{\psi }\dot{u} {}_{(\mu} u {}_{\nu)}
    + \frac{ \sqrt{2X} }{\psi }  \sigma  {}_{ \mu\nu} \nn\\
    &\nn\\
    &+ \frac{Z_{AB  } \nabla _{\mu}\phi ^{A } \nabla _{\nu}\phi ^{B  }}{\psi } - \frac{g_{ \mu\nu} Z_{A  B  } \nabla _{\alpha}\phi ^{A } \nabla ^{\alpha}\phi ^{B  }}{2 \psi } \,.
\end{align}
Using the decomposition $\nabla_{\mu}=h_{\mu}{}^{\nu}\nabla_{\nu}-u_{\mu} \, u^{\nu}\nabla_{\nu}$, defining $\dot{\phi}^{A}\equiv u^{\alpha}\nabla_{\alpha}\phi^{A}$, and taking into account the symmetry $Z_{AB}=Z_{BA}$, the interacting terms contribute to the density, pressure, heat flux and anisotropic stress,
\begin{align}
    Z_{AB}\nabla _{\mu}\phi ^{A } \nabla _{\nu}\phi ^{B  }=\,& Z_{AB}\Big(\dot{\phi}^{A}\dot{\phi}^{B}u_{\mu}u_{\nu}+ h_{\mu}{}^{\rho}h_{\nu}{}^{\sigma}\nabla_{\rho}\phi^{A}\nabla_{\sigma}\phi^{B}\nn\\
    &\nonumber\\
    &-2h_{(\mu}{}^{\alpha}u_{\nu)}\nabla_{\alpha}\phi^{A}\dot{\phi}^{B}\Big)
\end{align}
and the stress-energy tensor reads
\begin{align}
    \kappa \, T_{ \mu\nu}=\,&\Bigg(\frac{ V}{2 \psi }+ \frac{ X {} \omega }{\psi ^2} + \frac{\sqrt{2X}}{ \psi } \, \Theta+ \frac{Z_{AB}}{\psi} \, \dot{\phi}^{A}\dot{\phi}^{B}\nn\\
    &\nn\\
    &+\frac{Z_{AB}}{2\psi} \, \nabla _{\alpha} \phi ^{A } \nabla ^{\alpha}\phi ^{B  }\Bigg)u_{\mu}u_{\nu}\nn\\
    &\nn\\
    &+\Bigg(- \frac{ V}{2 \psi }+ \frac{ X {} \omega }{\psi ^2}- \frac{\dot{X} {}}{ \sqrt{2X} \psi }- \frac{2 \sqrt{2X}   }{3 \psi }\Theta  \nn\\
    &\nn\\
    &-\frac{Z_{AB}}{2\psi} \, \nabla _{\alpha}\phi ^{A } \nabla ^{\alpha}\phi ^{B  }\Bigg)h_{ \mu\nu}\nn\\
    &\nn\\
    &- \frac{2\sqrt{2X}  }{\psi } \, \dot{u} {}_{(\mu} u {}_{\nu)}-2\frac{Z_{AB}}{\psi}h_{(\mu}{}^{\alpha}u_{\nu)}\nabla_{\alpha}\phi^{A}\dot{\phi}^{B}\nn\\
    &\nn\\
    &+ \frac{ \sqrt{2X} }{\psi }  \sigma  {}_{ \mu\nu} +\frac{Z_{AB}}{\psi} \, h_{\mu}{}^{\rho}h_{\nu}{}^{\sigma}\nabla_{\rho}\phi^{A}\nabla_{\sigma}\phi^{B} \,.
\end{align}
Then, it is straightforward to obtain the effective fluid quantities
\begin{align}
    \kappa\,\rho=\,&\frac{1}{2 \psi ^2}\Big[\psi\, V + 2 X {} \omega + 2 \psi\sqrt{2X  }\, \Theta   \nn\\
    &\nn\\
    &+  \psi\,Z_{AB }  \left(\nabla _{\alpha}\phi ^{A  } \nabla ^{\alpha}\phi ^{B  } + 2 \dot{\phi} ^{A  } \dot{\phi} ^{B  }\right)\Big]\,,\\
&\nn\\
    \kappa\,q^{\alpha}=\,&-h ^{\alpha}{}_{\mu}u _{\nu}T^{ \mu\nu}\nn\\
    &\nn\\
    =\,&- \frac{\sqrt{2X }}{\psi } \, \dot{u} {}^{\alpha}- \frac{ Z_{AB  }}{\psi }   \, \dot{\phi} ^{A  }\, h ^{\alpha}{}_{\mu}\nabla^{\mu}\phi ^{B  }\,,\\
&\nn\\
    \kappa\,P=\,&- \frac{ V}{2 \psi }+ \frac{ X {} \omega }{\psi ^2}- \frac{\dot{X} {}}{ \sqrt{2X} \psi }- \frac{2 \sqrt{2X}   }{3 \psi } \, \Theta  \nn\\
    &\nn\\
    &-\frac{Z_{AB}}{2\psi} \, \nabla _{\alpha}\phi ^{A } \nabla ^{\alpha}\phi ^{B  }+\frac{Z_{AB}}{3\psi}h_{ \mu\nu}\nabla^{\mu}\phi^{A}\nabla^{\nu}\phi^{b}\nn\\
    &\nn\\
    =\,&\frac{\sqrt{2X  }}{3 \psi } \, \Theta  - \frac{\kappa T^\mathrm{(m)}  }{ \psi  (3 + 2 \omega )}- \frac{ 
    ( 2 \omega -1 )}{2 \psi  (3 + 2 \omega )} \, V +\frac{  X {} }{ \psi ^2 } \, \omega\nn\\
    &\nn\\
    &- \frac{  (\partial _{\psi }{}V \psi  + 2 X {} \partial _{\psi }{}\omega )}{ \psi  (3 + 2 \omega )}\nn\\
    &\nn\\
    &+\frac{  Z_{A  B  } (3 -2 \omega )-6 \partial _{\psi }{}Z_{A  B  } \psi   }{6 \psi  (3 + 2 \omega )} \, \nabla _{\alpha}\phi ^{B  } \, \nabla ^{\alpha}\phi ^{A  }\nn\\
    &\nn\\
    &+\frac{2 \sqrt{2X  }\, }{ \psi  (3 + 2 \omega )} \, \partial_{A}{}\omega \, \dot{\phi} ^{A  }+\frac{ Z_{A  B  }}{3 \psi  } \, \dot{\phi} ^{A  } \dot{\phi} ^{B  }\,,\nn\\
&\nn\\
    \kappa \,\pi^{\rho\sigma}=\,&\frac{\sqrt{2X} }{\psi } \, \sigma ^{\rho\sigma}+\left(h^{\mu\rho}h^{\nu\sigma}-\frac{1}{3}h^{\rho\sigma}h^{ \mu\nu}\right)\frac{Z_{AB}}{\psi}\nabla_{\mu} \, \phi^{A}\nabla_{\nu}\phi^{B}\nn\\
     &\nn\\
     =\,&\frac{\sqrt{2X} }{\psi } \,  \sigma^{\rho\sigma}+\frac{Z_{AB } \nabla ^{\rho}\phi ^{A  } \nabla ^{\sigma}\phi^B }{\psi }-\frac{h^{\rho\sigma} Z_{AB } \nabla _{\alpha}\phi^A  \nabla ^{\alpha}\phi ^{B}}{3\psi }\nn\\
     &\nn\\
     &+\frac{2 Z_{AB  }\dot{\phi}^{B} u^{(\rho} \nabla ^{\sigma)}\phi ^{A  }}{\psi }-\frac{ (g^{\rho\sigma}  -2 u^{\rho} u^{\sigma}) Z_{AB} \dot{\phi} ^{A  } \dot{\phi}^{B}  }{3 \psi }\nn\\
     &\nn\\
     =\,&\frac{\sqrt{2X} }{\psi }\,  \sigma^{\rho\sigma}+\frac{Z_{AB } \nabla ^{\rho}\phi ^{A  } \nabla^{\sigma}\phi^B }{\psi }-\frac{h^{\rho\sigma} Z_{AB } \nabla _{\alpha}\phi^A  \nabla ^{\alpha}\phi ^{B}}{3\psi }\nn\\
     &\nn\\
     &+\frac{ 2Z_{AB  }\dot{\phi}^{B} u^{(\rho} h^{\sigma)\alpha} \nabla _{\alpha}\phi ^{A  }}{\psi }-\frac{ (g^{\rho\sigma} +4 u^{\rho} u^{\sigma}) Z_{AB} \dot{\phi} ^{A  } \dot{\phi}^{B}  }{3 \psi }\,,
\end{align}
where an overdot denotes differentiation along the lines of the $\psi$-fluid, {\it i.e.}, $\dot{\phi}^A\equiv u^{\alpha}\nabla_{\alpha}\phi^{A}$.\\

At this point, we can identify the various contributions to the effective energy tensor as
\begin{align}
    P=\,&P_\mathrm{inv}+P_\mathrm{vis}+P_\mathrm{\phi}\\
    &\nonumber\\
    =\,&P_\mathrm{inv}-\zeta\Theta+P_\mathrm{\phi} \,,\\
    &\nn\\
    \rho=\,&\rho_\mathrm{inv}+\rho_\mathrm{vis}+\rho_\mathrm{\phi}\\
    &\nn\\
    =\,&\rho_\mathrm{inv}-3\zeta\Theta+\rho_\mathrm{\phi} \,,\\
    &\nn\\
    q^{\mu}=\,&- \frac{\sqrt{2X }}{\psi}\dot{u}^{\mu}+q^{\mu}_\mathrm{\phi} \,,\\
    &\nn\\
    \pi^{\mu\nu}=\,&-2\eta\sigma^{\mu\nu}+\pi^{\mu\nu}_\mathrm{\phi} \,,
\end{align}
where
\begin{align}
    \kappa\,P_\mathrm{inv}=\,&\frac{  X {} }{ \psi ^2 } \, \omega- \frac{ 
    ( 2 \omega -1 )}{2 \psi  (3 + 2 \omega )} \, V - \frac{\kappa T^\mathrm{(m)}  }{ \psi  (3 + 2 \omega )} \nn\\
    &\nn\\
    &- \frac{  (\partial _{\psi }{}V \psi  + 2 X {} \partial _{\psi }{}\omega )}{ \psi  (3 + 2 \omega )} \,,\nn\\
&\nn\\
    \kappa\,P_\mathrm{\phi}=\,&\frac{  Z_{A  B  } (3 -2 \omega )-6 \partial _{\psi }{}Z_{A  B  } \psi   }{6 \psi  (3 + 2 \omega )}\nabla _{\alpha}\phi ^{B  } \nabla ^{\alpha}\phi ^{A  }\nn\\
    &\nn\\
    &+\frac{2 \sqrt{2X  }\, }{ \psi  (3 + 2 \omega )}\partial_{A}{}\omega \, \dot{\phi} ^{A  }+\frac{ Z_{A  B  }}{3 \psi  } \dot{\phi} ^{A  } \dot{\phi} ^{B  }\,,\nn\\
&\nn\\
    \kappa\,\rho_\mathrm{inv}=\,&\frac{1}{2 \psi ^2}\left(\psi\, V + 2 X {} \omega + 2 \psi\sqrt{2X  }\right)\,,\\
&\nn\\
    \kappa\,\rho_\mathrm{\phi}=\,&\frac{Z_{AB }}{2 \psi }  \left(\nabla _{\alpha}\phi ^{A  } \nabla ^{\alpha}\phi ^{B  } + 2 \dot{\phi} ^{A  } \dot{\phi} ^{B  }\right)\,,\\
&\nn\\
    \kappa\,q^{\mu}_{\phi}=\,&- \frac{ Z_{AB  }}{\psi }   \dot{\phi} ^{A  }\, h ^{\mu}{}_{\alpha}\nabla^{\alpha}\phi ^{B  }\,,\\
&\nn\\
    \kappa\,\pi^{\mu\nu}_{\phi}=\,&\left(h^{\mu\rho}h^{\nu\sigma}-\frac{1}{3}h^{\rho\sigma}h^{ \mu\nu}\right)\frac{Z_{AB}}{\psi}\nabla_{\rho}\phi^{A}\nabla_{\sigma}\phi^{B}\,,
\end{align}
while
\begin{align}
    \zeta=\,-\frac{ \sqrt{2X}}{3\kappa\,\psi}\,,\quad\quad \eta=\,\frac{ \sqrt{2X}}{2\kappa\,\psi} 
\end{align}
are the bulk and shear viscosity coefficients, respectively.

In this particular case in which the Lagrangian is linear in $X$, the $\psi$-equation of motion reveal that $\Box\psi$ does not contain derivatives of the $\psi$-fluid four-velocity, therefore it only contributes to the inviscid pressure. However, it contains $\phi$-terms related to the interactions.

Finally, the $\phi$-terms contribute only to the inviscid part of the effective stress-energy tensor (because $P_\phi$ and $\rho_\phi$ depend only on first derivatives of the fields), to the heat flux, and to the shear viscosity.
In the general case of the previous section, all the $\phi$ fields contribute to both viscous and inviscid part.

\section{Conclusions}
\label{sec:7}

The picture of the effective fluid equivalent of tensor-multi-scalar 
gravity that emerges from the 
previous sections is the following. Because all the $N$ original 
gravitational scalar fields couple explicitly to the Ricci scalar, they are 
automatically 
coupled to each other. In addition, they may have explicit couplings to 
each other through the functions $Z_{AB}$ and $V$, but this is not 
necessary for them to be mutually coupled. In the multi-fluid 
interpretation, this property could correspond to these fields 
being 
thermalized, but this interpretation is not corroborated in any 
obvious way by 
the field equations and remains rather arbitrary.

\begin{acknowledgements}

M.~M. is grateful for the support of Istituto Nazionale di Fisica Nucleare 
(INFN) iniziativa specificha MOONLIGHT2, and for the 
hospitality at Bishop's University. This work is supported, in part, by
the Natural Sciences \& Engineering Research Council of Canada (grant  
2016-03803 to V.~F.).

\end{acknowledgements}

\noindent {\small The authors declare no conflict of interest.}

\appendix
\section{Kinematic quantities of the $\psi$-fluid}
\label{sec:appendix}

The (double) projection of the velocity gradient onto the 3-space 
orthogonal to $u^c$  
\be
V_{\alpha\beta} \equiv  {h_\alpha}^{\mu} \, {h_{\beta}}^{\nu} \, 
\nabla_{\nu} u_{\mu}  \,, \label{Vab}
\ee 
is decomposed into its symmetric and 
antisymmetric parts. The latter is identically zero because the 
$\psi$-fluid is derived from a scalar field. The symmetric part is 
further decomposed into its trace-free and  pure trace parts. This results 
in 
\be
V_{\alpha \beta}=  \Theta_{\alpha\beta} +\omega_{\alpha \beta} 
=\sigma_{\alpha \beta} +\frac{\Theta}{3} \, 
h_{\alpha \beta}+ \omega_{\alpha \beta} \,,
\ee
where the expansion tensor $\Theta_{\alpha \beta}=V_{(\alpha \beta)}$  is 
the symmetric part of 
$V_{\alpha \beta}$, $\Theta\equiv {\Theta^{\rho}}_{\rho} =\nabla_{\rho} 
u^{\rho} $ is its trace, the 
vorticity tensor  $\omega_{\alpha \beta}=V_{[\alpha \beta]} = 0 $, and the 
symmetric, trace-free shear tensor is 
\be 
\sigma_{\alpha \beta} \equiv \Theta_{\alpha \beta}-\frac{\Theta}{3}\, 
h_{\alpha \beta } \,.
\ee
Expansion, vorticity, and shear are purely spatial,
\begin{eqnarray}
\Theta_{\alpha \beta}u^{\alpha} &=& \Theta_{\alpha \beta}u^{\beta} = 
\omega_{\alpha \beta} \, u^{\alpha} = \omega_{\alpha \beta} \, u^{\beta} 
=  \sigma_{\alpha \beta} u^{\alpha}\nonumber\\
&&\nonumber\\
& =&  \sigma_{\alpha \beta} 
u^{\beta} = 0 \,.
\end{eqnarray}
For general fluids, it is \cite{Ellis:1971pg,EMMacC}  
\begin{eqnarray}
\nabla_{\beta} u_{\alpha} &=&   
\sigma_{\alpha \beta}+\frac{\Theta}{3} \, h_{\alpha \beta} +\omega_{\alpha 
\beta}  -  \dot{u}_{\alpha} 
u_{\beta}  =V_{\alpha \beta} -\dot{u}_{\alpha} u_{\beta} \,.\nonumber\\
&&  
\label{ecce}
\end{eqnarray}
The kinematic quantities of the $\psi$-fluid relevant for the present 
discussion are computed in \cite{Faraoni:2018qdr} and are as 
follows:
\begin{eqnarray}
\nabla_{\beta} u_{\alpha} &=& \frac{1}{ \sqrt{ -\nabla^{\rho}\psi 
\nabla_{\rho} \psi}} \left( 
\nabla_{\alpha} \nabla_{\beta} \psi -\frac{  \nabla_{\alpha} \psi 
\nabla^{\rho}  \psi \nabla_{\beta} 
\nabla_{\rho} \psi}{\nabla^{\sigma} \psi \nabla_{\sigma} \psi} \right) 
\,,\nonumber\\
&&
\end{eqnarray}
the acceleration is 
\begin{eqnarray}
\dot{u}_{\alpha} & \equiv &  u^{\beta}\nabla_{\beta} u_{\alpha} = \left( 
-\nabla^{\rho}\psi \nabla_{\rho} \psi \right)^{-2} 
\nabla^{\beta} \psi \nonumber\\
&&\nonumber\\
&\, & 
\Big[ (-\nabla^{\rho} \psi  \nabla_{\rho} \psi)  \nabla_{\alpha} 
\nabla_{\beta} 
\psi + \nabla^{\rho}  \psi \nabla_{\beta} \nabla_{\rho} \psi 
\nabla_{\alpha} \psi \Big] \,, 
\label{acceleration}
\end{eqnarray}   

\begin{eqnarray}
V_{\alpha\beta} &=& \frac{ \nabla_{\alpha}  
\nabla_{\beta} \psi }{ \left( -\nabla^{\rho}  \psi \nabla_{\rho} \psi 
\right)^{1/2} }  
\nonumber\\
&&\nonumber\\
&\, & +\frac{ \left( \nabla_{\alpha} 
\psi \nabla_{\beta}  \nabla_{\sigma} \psi + \nabla_{\beta} \psi  
\nabla_{\alpha} \nabla_{\sigma} \psi 
\right) \nabla^{\sigma}  
\psi }{ \left( -\nabla^{\rho} \psi \nabla_{\rho}  \psi \right)^{3/2} } 
\nonumber\\
&&\nonumber\\
&\, &  + \frac{ \nabla_{\delta} \nabla_{\sigma} \psi 
\nabla^{\sigma} \psi \nabla^{\delta} \psi }{\left( -\nabla^{\rho} 
\psi \nabla_{\rho}  \psi \right)^{5/2} } \, \nabla_{\alpha} \psi 
\nabla_{\beta} \psi \,. 
\end{eqnarray}  
The expansion scalar  reads
\begin{eqnarray} 
\Theta = \nabla_{\rho} u^{\rho} &=& 
\frac{ \square  \psi}{ \left (-\nabla^{\rho} \psi 
\nabla_{\rho} \psi \right)^{1/2} } \nonumber\\
&&\nonumber\\
&\, &  + \frac{ \nabla_{\alpha}
\nabla_{\beta} \psi \nabla^{\alpha} \psi \nabla^{\beta} \psi }{ \left( 
-\nabla^{\rho} \psi 
\nabla_{\rho}  
\psi \right)^{3/2} } \,, \label{thetaScalar}
\end{eqnarray}
while the shear tensor is 
\begin{eqnarray}
 \sigma_{\alpha\beta} 
&=&  \left( -\nabla^{\rho} \psi \nabla_{\rho} \psi \right)^{-3/2} \left[ 
-\left( \nabla^{\rho}  
\psi \nabla_{\rho} 
\psi \right) \nabla_{\alpha} \nabla_{\beta} \psi \right.\nonumber\\
&&\nonumber\\
&\, & \left. 
- \frac{1}{3} \left(  \nabla_{\alpha} \psi \nabla_{\beta} \psi   - 
g_{\alpha\beta} \, \nabla^{\sigma}
\psi \nabla_{\sigma} \psi   \right) \square \psi  \right.\nonumber\\
 &&\nonumber\\ 
&\, &  - \frac{1}{3} \left( g_{\alpha\beta} + \frac{ 2 
\nabla_{\alpha}  \psi 
\nabla_{\beta} \psi }{   \nabla^{\rho} \psi \nabla_{\rho} \psi } 
 \right) \nabla_{\sigma} \nabla_{\tau}
\psi \nabla^{\tau} \psi \nabla^{\sigma} \psi \nonumber\\
&&\nonumber\\
&\, & \left.
+ \left( \nabla_\alpha \psi 
\nabla_{\sigma} \nabla_{\beta} \psi 
+ \nabla_{\beta} \psi \nabla_{\sigma} \nabla_{\alpha} \psi \right) 
\nabla^{\sigma} \psi \right]  \,.\label{shearofpsi}
\end{eqnarray}


\end{document}